\documentclass[manuscript, screen, authorversion]{acmart}

\AtBeginDocument{%
  }

\copyrightyear{2025}
\acmYear{2025}
\setcopyright{rightsretained}
\acmConference[AutomotiveUI Adjunct '25]{17th International Conference on Automotive User Interfaces and Interactive Vehicular Applications}{September 21--25, 2025}{Brisbane, QLD, Australia}
\acmBooktitle{17th International Conference on Automotive User Interfaces and Interactive Vehicular Applications (AutomotiveUI Adjunct '25), September 21--25, 2025, Brisbane, QLD, Australia}
\acmDOI{10.1145/3744335.3758480} 
\acmISBN{979-8-4007-2014-7/2025/09}

\acmSubmissionID{27}

\usepackage[T1]{fontenc}
\usepackage{xcolor}
\usepackage{xspace}
\usepackage[capitalise]{cleveref}
\usepackage{subcaption,amsfonts,dcolumn} 

\newcommand{\ie}{\emph{i.e.}\xspace}
\newcommand{\eg}{\emph{e.g.}\xspace}

\newcommand{\aka}{\emph{a.k.a.}\xspace}

\definecolor{Green}{RGB}{0,128,0}

\newtheorem*{definition}{Definition}
\definecolor{myorange}{RGB}{241, 109, 28}
\definecolor{mygreen}{RGB}{111, 138, 32}
\definecolor{myblue}{RGB}{46, 139, 159}
\definecolor{myred}{RGB}{219, 60, 86}
\definecolor{mydarkorange}{RGB}{172, 79, 23}
\definecolor{mydarkgreen}{RGB}{40, 93, 52}
\definecolor{mydarkblue}{RGB}{18, 66, 88}
\newcommand{\driverInvolvementIndex}{\textcolor{mydarkorange}{\text{DII}}\xspace}
\newcommand{\driverInvolvementDeficit}{\textcolor{myorange}{\text{DID}}\xspace}
\newcommand{\vehicleEngagementIndex}{\textcolor{myblue}{\text{VEI}}\xspace}
\newcommand{\vehicleEngagementDeficit}{\textcolor{mydarkblue}{\text{VED}}\xspace}
\newcommand{\environmentComplexityIndex}{\textcolor{mygreen}{\text{ECI}}\xspace}

\newcommand{\cumulativeDeficit}{\text{CD}\xspace}

\begin{document}

\title[DEV: A Driver-Environment-Vehicle Closed-Loop Framework]{DEV: A Driver-Environment-Vehicle Closed-Loop Framework for Risk-Aware Adaptive Automation of Driving} 

\author{Ana{\"i}s Halin}
\email{anais.halin@uliege.be}
\orcid{0000-0003-3743-2969}
\affiliation{%
  \institution{University of Liège}
  \department{Department of Electrical Engineering and Computer Science}
  \city{Liège}
  \country{Belgium}
}

\author{Christel Devue}
\email{cdevue@uliege.be}
\orcid{0000-0001-7349-226X}
\affiliation{%
  \institution{University of Liège}
  \department{Department of Psychology}
  \city{Liège}
  \country{Belgium}
}

\author{Marc Van Droogenbroeck}
\email{m.vandroogenbroeck@uliege.be}
\orcid{0000-0001-6260-6487}
\affiliation{%
  \institution{University of Liège}
  \department{Department of Electrical Engineering and Computer Science}
  \city{Liège}
  \country{Belgium}
}


\begin{abstract}
The increasing integration of automation in vehicles aims to enhance both safety and comfort, but it also introduces new risks, including driver disengagement, reduced situation awareness, and mode confusion.
In this work, we propose the \emph{DEV} framework, a closed-loop framework for risk-aware adaptive driving automation that captures the dynamic interplay between the driver, the environment, and the vehicle. 
The framework promotes to continuously adjusting the operational level of automation based on a risk management strategy. The real-time risk assessment supports smoother transitions and effective cooperation between the driver and the automation system. 
Furthermore, we introduce a nomenclature of indexes corresponding to each core component, namely driver involvement, environment complexity, and vehicle engagement, and discuss how their interaction influences driving risk. The \emph{DEV} framework offers a comprehensive perspective to align multidisciplinary research efforts and guide the development of dynamic, risk-aware driving automation systems.
\end{abstract}



\maketitle

\section{Introduction}

Vehicles are becoming increasingly automated, so that driving has truly become a cooperation between the human driver and the machine driver, commonly referred to as the driving automation system.
But for this cooperation to be truly effective and to take place under optimal conditions, we need a human-machine interface that is well designed~\cite{Naujoks2019Towards,Wang2024HumanMachine}, that is, an interface that allows drivers to clearly understand their responsibilities at all times, to form an accurate mental model, and to grasp what the driving automation system is doing and why it is doing it. Only then can drivers trust the system and play their role in the best conditions.

Currently, there are many issues where drivers lack awareness of automation limitations~\cite{Beggiato2013TheEvolution}, overtrust the system (\eg, caused by phenomena like autonowashing~\cite{Dixon2020Autonowashing}), or have mode confusion~\cite{Wilson2020Driver}. On the other hand, some drivers do not benefit from the advantages of driving automation at all because they never activate the available driving automation features~\cite{Orlovska2020Effects}.

The SAE levels~\cite{Sae2021Taxonomy} of driving automation do little to help drivers better understand their responsibilities or to support effective real-time cooperation between the human and the automation in driving~\cite{Inagaki2019ACritique,Boyle2024Workshop,Huang2025Beyond-arxiv}. At best, the SAE levels serve legal or manufacturer-related purposes.

Even if we assume the existence of an ideal communication interface between the human and the machine, one that enables perfect cooperation between the two, the question remains: how should the human and the machine cooperate to mitigate the risk? Drivers have limited capacities that fluctuate depending on their state and their willingness to engage in the driving task. 
Similarly, the vehicle is equipped with a driving automation system that includes more or fewer automation features (depending on its SAE level), which can be activated or deactivated. 
The driver state and the availability or appropriateness of automation both depend on the driving environment. For example, the literature tells us that a monotonous environment increases driver fatigue~\cite{Farahmand2018Effect}, while a complex environment leads to cognitive overload~\cite{Teh2014Temporal}. 
The vehicle, for its part, operates within an operational design domain (ODD) that restricts when and where automation features can be used~\cite{Sae2021Taxonomy}. Additionally, some driving automation features reduce the driver's mental workload by helping with certain tasks, sometimes even too much, to the point where the driver disengages entirely from the driving task~\cite{DeWinter2014Effects, Bai2025Awakening}. Others (such as warning systems) may actually increase the driver workload~\cite{Koniakowsky2024From}.

In short, there is a highly complex interaction between the driver, the driving automation system, and the driving environment. Understanding the full scope of this interaction is crucial to defining the optimal cooperation between the human and the automation system, one that enables safe driving and minimizes risk.

We therefore introduce a closed-loop framework that accounts for the driver, the vehicle, and the environment, and that enables dynamic adaptation of the operational level of automation to mitigate risk.
We discuss how the driver, the vehicle, and the environment each influence risk. 
This framework highlights the key questions that still need to be addressed to implement such a system.

\begin{figure*}
    \centering
    \includegraphics[width=\linewidth]{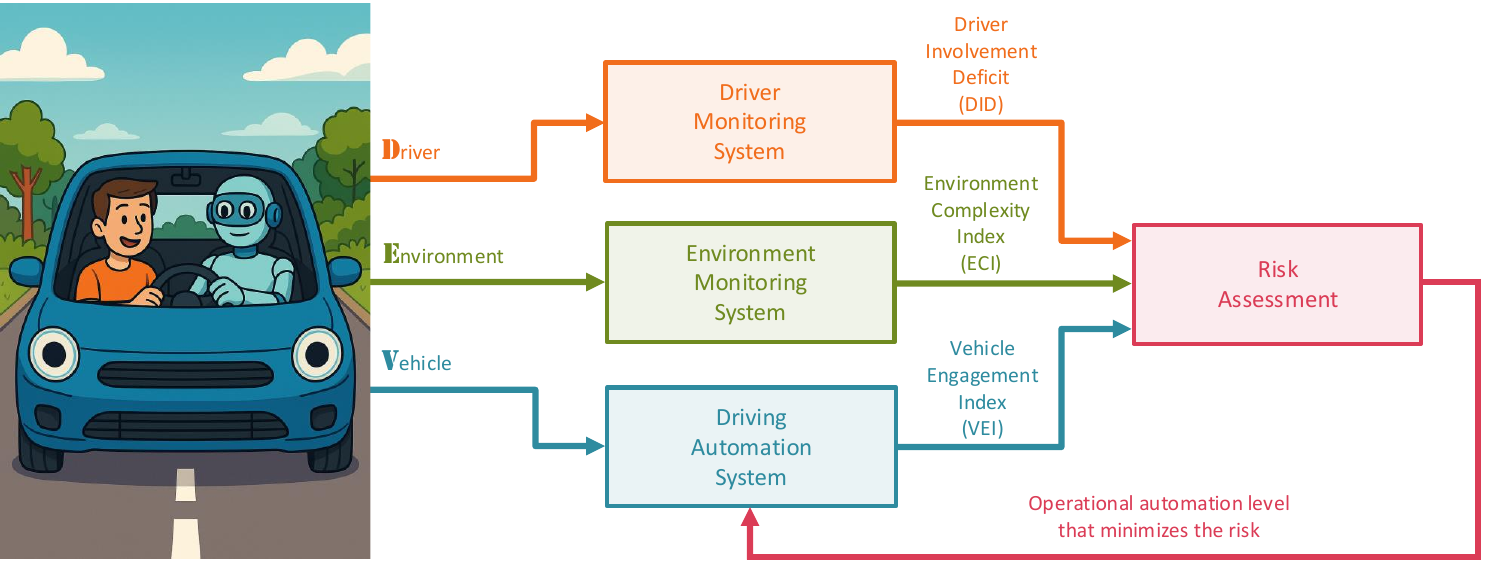}
    \Description{This is a block diagram depicting how the risk-aware adaptive automation is performed based on the DEV framework.}
    \caption{\textbf{The DEV closed-loop framework for risk-aware adaptive automation.} 
    At all SAE levels except level 5, the driver and the vehicle collaborate to perform the driving task. The operational automation level of the driving automation system can be dynamically adjusted to mitigate the risk by accounting for three core dimensions: the driver, the environment, and the vehicle. 
    A driver involvement deficit (\driverInvolvementDeficit) is computed using a driver monitoring system, based on the driver state and fitness to drive. An environment complexity index (\environmentComplexityIndex) is derived from an environment monitoring system using data such as traffic conditions, road type, and visibility. The vehicle engagement index (\vehicleEngagementIndex) reflects the extent to which the vehicle via the driving automation system is involved in the driving task.
    Risk is assessed based on \driverInvolvementDeficit, \environmentComplexityIndex, and \vehicleEngagementIndex, and the system adjusts the operational level of automation accordingly. 
    }
    \label{fig:DEV_framework}
\end{figure*}

\section{The DEV Closed-Loop Framework for Risk-Aware Adaptive Automation}

In this section, we introduce the DEV closed-loop framework, which supports risk-aware adaptive automation by dynamically adjusting the operational level of vehicle automation (see \cref{fig:DEV_framework}). This adjustment is based on three core components: (1) driver involvement, (2) environment complexity, and (3) vehicle engagement. We define these components below and give some leads to assess risk to guide the adaptation process.

\subsection{Driver Involvement}
\begin{definition}
    \emph{Driver involvement} is defined as the proportion of driver (cognitive, perceptual, and physical) resources that are available to perform the driving task.
\end{definition}

High driver involvement indicates that the driver is fully fit and ready to drive, while low driver involvement suggests that the driver is in a state where taking control of the vehicle would be undesirable or unsafe.
Driver involvement depends on the driver state (\eg, drowsiness, distraction) and, to some extent, on their willingness to engage in the driving task. For instance, a driver may be fully awake but choose to disengage out of laziness or disinterest, thereby reducing their effective involvement.

While driver involvement cannot be measured directly, it can be estimated. We define two complementary numbers, both ranging from $0$ to $1$: the driver involvement index (\driverInvolvementIndex) and its complement, the driver involvement deficit (\driverInvolvementDeficit), such that $\driverInvolvementDeficit=1-\driverInvolvementIndex$. 
A driver involvement index of $1$ (\ie, a deficit of $0$) indicates that the driver is fully capable of performing the entire driving task. Conversely, an index of $0$ (\ie, a deficit of $1$) indicates that the driver is entirely unable to contribute to the driving task. Note that an index or a deficit should have various properties such as being measurable, increasing (or decreasing), on a scale (which can be discrete), and reproducible.

The driver involvement deficit \driverInvolvementDeficit can be estimated by assessing the various components affecting the driver state, such as drowsiness, mental workload, distraction, emotions, and under influence~\cite{Halin2021Survey}. One possible approach is to compute a weighted average of these components to obtain an overall estimate. Another approach is to directly assess a global estimate of the driver involvement deficit through the implementation of the Minimum Required Attention (MiRA) theory~\cite{Kircher2016Minimum, Ahlstrom2022Towards}. According to MiRA, drivers are considered attentive if they sample sufficient information from the environment to meet the demands of the driving task, or, in other words, if they fulfill the preconditions needed to form and maintain a good enough mental representation of the situation. Drivers are thus considered inattentive only when information sampling is not sufficient, regardless of whether they are drowsy, concurrently executing an additional task or not.

\subsection{Environment Complexity}
\begin{definition}
    \emph{Environment complexity} is defined as the proportion of (cognitive, perceptual, and physical) resources that the driver would need to safely perform the driving task in a given context.
\end{definition}

The more complex the environment, the greater the demands placed on the driver, making the driving task more challenging. Environment complexity is conceptualized from the perspective of the driver and should ideally reflect perceptual and cognitive factors, such as visual masking (\ie, the difficulty in perceiving a target stimulus when it is rapidly followed or spatially overlapped by another stimulus) or joint attention with other road users (\ie, the coordination of attentional focus between the driver and other road users, which plays a critical role in interpreting intentions and ensuring safe interactions)~\cite{Rasouli2020Autonomous,Kotseruba2020Joint}.

For the environment complexity, we also define an index, named environment complexity index  (\environmentComplexityIndex), which ranges from $0$ (minimal complexity) to $1$ (maximal complexity).

The environment complexity depends not only on the number of elements present, but also on their behavioral relevance~\cite{Yu2021Dynamic}. Simply counting objects in the driving scene is not sufficient to estimate the environment complexity index. For instance, a group of pedestrians conversing on the sidewalk may contribute less to complexity than a single pedestrian approaching a crosswalk.
Other factors, such as road type, road geometry, and weather conditions, also play a role in shaping the complexity of the driving environment. 
However, how to accurately compute the environment complexity index in a way that is meaningful for risk assessment remains an open question that current research has largely overlooked.

\subsection{Vehicle Engagement}
\begin{definition}
    \emph{Vehicle engagement} is defined as the proportion of vehicle automation resources that are available to perform the driving task. 
\end{definition}

Vehicle engagement reflects the set of driving automation features that can be activated to assist the driver in performing the driving task. It varies depending on the driving environment and the ODD, which defines the limitation of the driving automation system. For instance, some features could only be activated on highways and clear weather conditions (avoiding adverse conditions like heavy rain). Besides, the maximum achievable vehicle engagement is actually limited by the SAE level of the vehicle. 

To quantify this, we define the vehicle engagement index (\vehicleEngagementIndex), which ranges from $0$ (no automation) to $1$ (full automation). We also define its complement, the vehicle engagement deficit (\vehicleEngagementDeficit) as $\vehicleEngagementDeficit=1-\vehicleEngagementIndex$.

Vehicle engagement depends on the availability of driving automation features such as (adaptive) cruise control, lane centering assistance, lane keeping assistance, automatic emergency braking, rear automatic braking, pedestrian automatic emergency braking, blind spot intervention, and others. The scale is absolute, with 
$\vehicleEngagementIndex=1$ denoting full automation, attainable exclusively by a SAE Level 5 vehicle. Using an absolute rather than a relative scale facilitates interpretation and comparison between vehicles. 

\subsection{Risk Assessment}
\begin{definition}
    In the context of our framework, \emph{risk} refers to the likelihood of adverse outcomes, such as collisions, traffic violations or even near-misses, resulting from the interaction between the driver, the environment, and the vehicle.
\end{definition}

Risk results from the real-time ability (or inability) to safely perform the driving task. 
It is influenced by the driver involvement, the environment complexity, the vehicle engagement, and by how they interact (\eg, the nature of the cooperation between the human driver and driving automation system of the vehicle, but also the design of the human-machine interface). 
    
The goal of risk assessment is to enable dynamic adjustment of the operational level of automation by activating or deactivating specific driving automation features in a closed-loop fashion, thereby mitigating the computed risk. By continuously adapting the level of automation based on driver involvement, environmental complexity, and vehicle engagement, we aim to support smooth transitions of control between the human driver and the automation system. 

Such gradual transitions offer several advantages. First, they allow the driver to progressively adapt their mental model, avoiding abrupt changes in system behavior. Second, they help maintain an appropriate level of driver involvement relative to the demands of the current driving context and associated risk.

However, assessing risk based on the intricate interplay between the driver, the environment, and the vehicle remains a major challenge. This requires a deep understanding of how these three components influence one another. For instance, how specific driving automation features affect driver state under varying environment complexities. The work in~\cite{Halin2025Effects} takes a first step in this direction by analyzing the influence of cognitive load and traffic conditions on the use of an adaptive cruise control, and how its use in turn impacts driving performance.

To begin formalizing this interplay, we first consider a simplified case in which the environment complexity is held constant. This allows us to focus on how the relationship between driver and vehicle resources contributes to risk. Specifically, we examine the two-dimensional space defined by the vehicle engagement deficit (\vehicleEngagementDeficit) and the driver involvement deficit (\driverInvolvementDeficit). 
We define the cumulative deficit ($\cumulativeDeficit$) by $\cumulativeDeficit = \vehicleEngagementDeficit + \driverInvolvementDeficit -1$ such that $\cumulativeDeficit = 1$ when $\driverInvolvementDeficit=\vehicleEngagementDeficit=1$, $\cumulativeDeficit = 0$ when $\driverInvolvementDeficit=\vehicleEngagementDeficit=0.5$, and  $\cumulativeDeficit = -1$ when $\driverInvolvementDeficit=\vehicleEngagementDeficit=0$. 

\begin{figure}
    \centering
    \includegraphics[width=.7\linewidth]{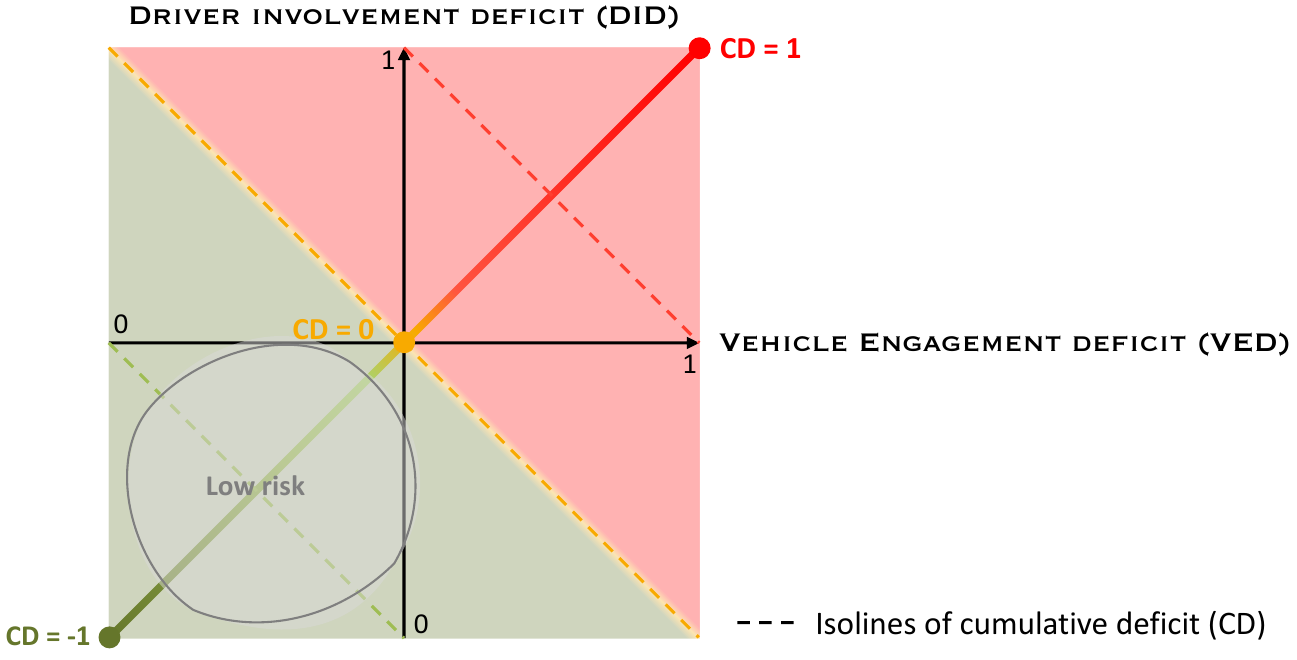}
    \Description{This figure shows how the cumulative deficit evolves on a graph as a function of the vehicle engagement deficit ranging from zero to one on the horizontal axis and the driver involvement deficit also ranging from zero to one on the vertical axis.}
    \caption{\textbf{Cumulative deficit in the two-dimensional space defined by \vehicleEngagementDeficit and \driverInvolvementDeficit, for a fixed \environmentComplexityIndex.} 
    The main diagonal (top-left to bottom-right) and its parallels (dashed lines) are isolines of the cumulative deficit, meaning that \cumulativeDeficit remains constant along each of these lines. The cumulative deficit ranges from $-1$ to $1$ along the direction of the minor diagonal (solid line running from the bottom-left corner to the top-right corner). The main diagonal delineates two regions: one where the cumulative deficit is negative (green area), indicating redundancy in the combined resources of the driver and the vehicle to perform the driving task; and one where it is positive (red area), indicating that the combination of the driver and vehicle resources is insufficient to fully carry out the driving task. The objective of the DEV framework is to manage the risk through adaptive automation to remain in a low-risk area (in gray-tinted green).}
    \label{fig:cumulative_deficit}
\end{figure}

\Cref{fig:cumulative_deficit} illustrates the two-dimensional space defined by \vehicleEngagementDeficit and \driverInvolvementDeficit, and shows how the cumulative deficit evolves within this space. By definition, the cumulative deficit ranges from $-1$ to $1$ along the minor diagonal. In contrast, in the direction of the main diagonal, the space is structured by isolines, each corresponding to a constant value of the cumulative deficit. On the main diagonal, the cumulative deficit is zero, meaning that combining the resources of the driver and the vehicle barely allows performing the entire driving task, assuming perfect cooperation between the two agents.
The main diagonal separates the space into two regions: one where the cumulative deficit is positive, and one where it is negative.
In the positive cumulative deficit region, the combined resources of the driver and the vehicle are insufficient to carry out the full driving task. This region is thus inherently associated with high risk and should be avoided by any means. As this region is approached, proactive measures must be taken to prevent accidents. One such measure is for the vehicle to initiate a safety stop, similar to SAE level 4 vehicles, which achieve a minimal risk condition automatically when reaching their ODD limits.
In the negative cumulative deficit region, there is redundancy. Combining the available resources of both the driver and the vehicle provides more resources than necessary to perform the driving task. In this region, there is a real interest in selecting the most appropriate operational level of automation to foster effective cooperation between the driver and the vehicle to lower the risk. 
Below the minor diagonal, where the vehicle engagement deficit is higher than the driver involvement deficit, the driver is able to perform a higher proportion of the driving task than the driving automation system of the vehicle. Conversely, above the minor diagonal, the driving automation system of the vehicle is able to perform a higher proportion of the driving task than the driver.

When in this negative cumulative deficit region, effective cooperation between the driver and the automation is expected. However, how this cooperation should be operationalized and which operational level of automation (\ie, which combination of driving automation features) should be selected to minimize risk remains an open question.
Several authors (\eg, \cite{Young2023ToAutomate,Endsley2016From}) suggest that the driver should be utilized at their highest level of involvement. If the driver is in a suitable state and willing to engage in the driving task, they should be allowed to do so. However, when driving automation features are available, they should also be leveraged to reduce risk. Caution is needed, though: increasing the level of automation may negatively impact driver involvement.
The operational level of automation can be selected within the range $[0,\vehicleEngagementIndex]$. This choice should account not only for the driver's current level of involvement but also for the complexity of the driving environment, so that the selected operational level of automation mitigates risk.
In low-complexity driving environments, allowing the driver to be less involved and delegating more responsibility to the vehicle (\ie, increasing the operational level of automation) may not significantly increase risk. However, as the complexity of the driving environment rises, greater driver involvement may become necessary. 

Finally, it is important to note that assessing risk and adjusting the operational level of automation cannot be done without also considering the mode of cooperation between the human driver and the driving automation system. Cooperation can take the form of traded control (\ie, working on the same task at different times~\cite{Sheridan1978Human}; \aka control shifts~\cite{Walch2019Driving}), shared control (\ie, working on the same task at the same time~\cite{Sheridan1978Human}), for example through haptic feedback mechanisms~\cite{Griffiths2005Sharing,Abbink2012Haptic}, or a combination of both. A recent taxonomy for human-vehicle cooperation has been proposed to aid in structuring cooperative control concepts~\cite{Peintner2024HowTo}. 
According to~\citet{deWinter2022Shared}, whether shared control reduces risk depends on the environment complexity. 
This highlights the importance of integrating considerations about cooperation modes into the risk assessment process of our DEV framework. However, how to incorporate different cooperation modes effectively remains an open area for future research.

\section{Discussion and Future Work}

Researchers have long studied human-machine interaction, yet many challenges remain, particularly in the automotive domain. Interest in this area continues to grow, driven by the complexity of designing systems that must integrate insights from multiple disciplines. This includes, for example, developing driver monitoring systems that meaningfully support risk mitigation, or designing human-machine interfaces that enable effective cooperation. A wide variety of studies are being conducted across these domains, but building fully functional systems requires a shared understanding and coordinated direction. This is why a framework that captures the big picture is crucial, as it can help guide research efforts and align them toward a common goal.

In this work, we proposed the DEV framework for risk-aware adaptive automation of driving, which aims to dynamically and continuously adjust the operational level of automation based on the complex interplay between the driver, the surrounding environment, and the vehicle.
This framework is motivated by the observation that the static definition of SAE levels is inadequate for real-time driving, as the level of automation may need to change during a single drive. 
Moreover, operational levels should be more fine-grained to enable smooth, dynamic transitions, depending, for example, on the driver’s level of involvement, and to account for different cooperation modes between the driver and the automation system.

To move beyond a purely theoretical proposal, several key research directions must be pursued. The first step is to refine the definitions and develop methods for estimating the driver involvement deficit, the environment complexity index, and the vehicle engagement index. While these parameters cannot be directly measured, they can be estimated. Developing reliable and interpretable estimation methods for each component is a prerequisite for implementation.

The next step is to empirically establish a model for risk assessment, either through the theoretical development of a mathematical formulation or by leveraging data-driven approaches, such as deep learning. 
For example, \citet{Herzberger2022Confidence} presented the concept of confidence horizons, aiming at quickly identifying whether safety-critical transitions of control (takeovers or handovers) should be performed. 
Then, \citet{Herzberger2024Cooperation} presented the theoretical concept of the diagnostic takeover request, which predicts risky takeovers based on a driver’s response to a takeover request.
These works are important first steps toward cooperative driving, but currently only allow to gain more reaction time during critical transitions.

Once the risk model is defined, it becomes possible to design, validate, and implement strategies to mitigate that risk (\ie, remain in the gray-tinted green area in~\cref{fig:cumulative_deficit}).
Within the DEV framework, we propose modulating the operational level of automation, by activating or deactivating driving automation features, as the primary strategy for risk mitigation. However, interventions can also target the driver or the environment. Actions on the driver, such as reducing distraction or increasing situation awareness (\eg, \cite{Wu2023Dangerous,Goodge2024CanYou}), should only be implemented when adjusting the automation level alone cannot adequately compensate for low driver involvement. Drivers should retain the freedom to rest or disengage from the driving task, so long as doing so does not significantly increase the risk.

Actions on the environment to reduce its complexity may include, \eg, improving road infrastructure and design, for instance, following the principles of \emph{Self-Explaining Roads}~\cite{Theeuwes1995Self}, which aim to encourage safe behavior through intuitive road layouts. However, such interventions can only be implemented occasionally and require substantial time and financial resources.

Finally, once such a system is implemented, its acceptability and usability must be thoroughly investigated. These aspects hinge largely on the design of the human-machine interface, which should support efficient cooperation between the driver and the vehicle. Drivers must clearly understand their role in the human-machine partnership, and be able to form accurate mental models of how the system operates and why it adapts in real time. Although recent work has addressed these challenges~\cite{Dixon2025Explaining,Weiser2025From}, much remains to be explored.

\begin{acks}
The work by A. Halin was supported by the SPW EER, Wallonia, Belgium under grant n°2010235 (ARIAC by \href{https://www.digitalwallonia.be/en/}{DIGITALWALLONIA4.AI}). 
\end{acks}

\bibliographystyle{ACM-Reference-Format}
\bibliography{main}

\end{document}